\documentclass[oneside]{IEEEtran}
\usepackage[cmex10]{amsmath}
\usepackage{amssymb}
\usepackage{cite}
\usepackage{amsmath}
\usepackage{amssymb}
\usepackage{booktabs}
\usepackage{balance}
\usepackage{subfigure}
\usepackage{multicol}
\usepackage{graphicx}
\usepackage{epsfig}
\usepackage{epstopdf}
\usepackage{algorithm}
\usepackage{algorithmic}
\usepackage{bm}
\usepackage{amsmath, amssymb, amsthm}
\usepackage{array}
\usepackage{makecell} \usepackage{slashbox}
\usepackage{multirow}% ±í¸ñ
\usepackage{float}% ÓÃÓڹ̶¨Í¼Æ¬Î»ÖÃ
\usepackage{multicol,multicap,graphicx}
\usepackage{1-in-2}
\usepackage{mathrsfs}
\usepackage{url}

\usepackage{ifpdf}
\usepackage[bf]{caption2}

\usepackage{color}

\ifCLASSINFOpdf
\else
\fi
\begin{document}
\title{\sc{NOMA Assisted Multi-MEC Offloading for IoVT Networks}}

% \author{\IEEEauthorblockN{}
 %\IEEEauthorblockA{~The Information Network Lab of EEIS Department USTC, Hefei, China, 230027 \\}
\author{Fengqian~Guo,
        Hancheng~Lu,
        Bo Li,
        Dingxuan~Li,
        and Chang Wen Chen
}

\maketitle

%Telephone: (800) 555--1212, Fax: (888) 555--1212}


%To support more IoVT devices, more BSs with MEC server are required.

\section*{\large{Abstract}}
Nowadays, Internet of Video Things (IoVT) grows rapidly in terms of quantity and computation demands. In spite of the higher local computation capability on visual processing compared with conventional Internet of Things devices, IoVT devices need to offload partial visual processing tasks to the mobile edge computing (MEC) server wirelessly due to its larger computation demands. However, visual processing task offloading is limited by uplink throughput and computation capability of the MEC server. To break through these limitations, a novel non-orthogonal multiple access (NOMA) assisted IoVT framework with multiple MEC servers is proposed, where NOMA is exploited to improve uplink throughput and MEC servers are co-located with base stations to provide enough computation capability for offloading. In the proposed framework, the association strategy, uplink visual data transmission assisted by NOMA and division of the visual processing tasks as well as computation resource allocation at the MEC servers are jointly optimized to minimize the total delay of all visual processing tasks, while meeting the delay requirements of all IoVT devices. Simulation results demonstrate that significant performance gains can be achieved by proposed joint optimization with NOMA transmission and multi-MEC offloading in the heterogeneous IoVT network.

\IEEEpeerreviewmaketitle
\mathfootnote{\centering\footnotesize{Fengqian~Guo, Hancheng~Lu, Bo Li and Dingxuan~Li are with University of Science and Technology of China;
Chang Wen Chen is with the Department of Computing, Hong Kong Polytechnic University, Hung Hom, Kowloon, Hong Kong SAR, China.}}

\section*{\large{Introduction}}

%The Internet of Things (IoT) technology has developed rapidly with billions of connected sensors deployed everywhere\cite{Xiang2020NOMA-Assisted,Feng2020UAV-Enabled,Elkashlan2019,Zhang2019b}. Among them, visual sensors (e.g cameras in smartphones, vehicles and buildings) have composed a new sub-field of IoT called the internet of video things (IoVT). Recently, these IoVT devices play a more and more important role in public security, traffic analysis, industrial automation and smart city \cite{Mohan2017,Acharya2019,Yang2017,Perala2018,Ji2020}.

The Internet of Things (IoT) technology is developing rapidly. Billions of connected sensors have been deployed for various applications. Among them, IoT with visual sensors (e.g, cameras in smartphones, vehicles and buildings) have emerged into a new sub-field of IoT called Internet of Video Things (IoVT)\cite{Chen2020}{, or visual IoT\cite{Ji2019Visual}}. With unique characteristics in terms of sensing, transmission, storage, and analysis, IoVT has gained much attention from academia and industry. It has been applied in specific scenarios like public security, traffic analysis, industrial automation and smart city \cite{Perala2018,Ji2020Crowd,Chen2016,Guiyongqiang2020MC}, which imposes high requirements on computation and transmission resources.

The most distinctive feature of IoVT lies in processing of sensored visual data. Most IoVT applications {or visual IoT applications} rely on visual processing to achieve specific objectives, for example, recognition and tracking. {In \cite{Xiong2019a}, to analyze the human action in sport training videos, a deep learning based action recognition method is proposed .
The proposed method utilizes mathematical models like neural networks and image processing techniques to sense the environment. A compact deep neural network-based face recognition method for face recognition in public safety surveillance system with visual IoT is presented in \cite{Oh2018}.
A novel visual IoT architecture considering the interaction among visual IoT devices is proposed in \cite{Ji2019Visual} to improve end-to-end performance for next generation smart cities.} In \cite{Blanco-Filgueira2019}, low-power and real-time multiple object visual tracking is implemented in an IoVT network with a camera and wireless connection capability.

Different from conventional IoT devices, in general, IoVT devices have higher local computation capability for visual processing \cite{Chen2020}. In spite of this, visual processing task offloading is still necessary in IoVT networks, due to the huge computation requirements on visual processing that cannot be met by IoVT devices. It has been proposed to offload visual processing tasks to the mobile edge computing (MEC) server with tolerant delay \cite{Ji2020Crowd,Perala2018}. To implement MEC offloading, MEC servers are deployed at the edge of radio access networks (e.g., base station (BS)). Limited by the cost and hardware factors, IoVT devices may not be able to finish all of their visual processing tasks by themselves on time. In this case, IoVT devices can offload part of their visual processing tasks to MEC servers \cite{Chen2020}. Usually, IoVT devices tend to handle the visual processing tasks with higher delay sensitive and lower computation demands locally, and offload the visual processing tasks with lower delay sensitive and higher computation demands to MEC servers.

Many research attempts have been done on MEC offloading in IoVT networks, where delay and energy consumption are mostly concerned. In \cite{Wang2019b}, tremendous visual data is reduced by deep learning models in computational IoVT devices and the transmission function is optimized for high efficiency, with the goal to obtain the least transmission delay for MEC offloading under dynamic uplink bandwidth. To reduce system energy consumption considering the resource constraints of the IoVT devices and MEC server, authors in \cite{Perala2018} optimize the content generation rates of IoVT devices to make full use of the computation resources at the MEC server. For urban traffic surveillance characterized by low delay and real-time processing, the dynamic video stream processing scheme is investigated in \cite{Chen2016}, aiming at reducing the processing time at the MEC server. In \cite{Trinh2018}, the MEC offloading policy is optimized by differential treating low-to-high workloads of visual data processing so that the IoVT requirements on energy consumption and processing delay can be satisfied. Particularly, the MEC server gives priority to the IoVT device desiring energy conservation over low latency with lower workload. To meet the delay and energy constraints in the IoVT network, authors in \cite{Sultana2019} present a generic architecture of an IoVT based video surveillance system, where the image-video compression and coding tasks are offloaded to the MEC server. {In order to satisfy the low latency and low energy consumption requirements of visual IoT devices, the author of \cite{Ji2020Crowd} propose a visual IoT architecture for MEC offloading, where the conventional multimedia streaming is changed to feature-type transmission.}

However, aforementioned work only considers single MEC server offloading in IoVT networks and might not provide enough computation resources to meet the visual processing task offloading demands. Compared with traditional IoT, IoVT produces a much larger amount of data for processing. Moreover, processing of visual data is much more complex than that of common IoT data. In the context of MEC offloading, it means that IoVT imposes much more requirements on computation resources. This observation motivates us to consider the IoVT scenario where multiple MEC servers are deployed with BSs and investigate multi-MEC offloading. Another issue is the limitation of wireless uplink throughput for visual data transmission. Most existing work uses compression technologies to reduce the amount of transmitted visual data, leading to additional pressure on computation resources and energy consumption at the IoVT devices. Differently, we introduce non-orthogonal multiple access (NOMA) \cite{Wu2018Noma,ZhangMing2021MC} in the IoVT network to achieve more wireless uplink throughput. Particularly, power allocation is optimized with consideration of the power constraints at the IoVT devices.

In this article, we propose a novel NOMA assisted IoVT framework with multi-MEC. In the proposed framework, considering the heterogeneity of wireless channel conditions as well as computation capability at IoVT devices and MEC servers, we perform jointly optimization on the association strategy, the power allocation strategy of uplink NOMA, division of visual processing tasks and the way that MEC servers allocate their computation resources, to minimize the total delay of all visual processing tasks while meeting the delay requirements of all IoVT devices.

The rest of this article is organized as follows. An IoVT network with multi-MEC as well as the challenges for multi-MEC offloading are firstly described. Then, to address these challenges, a novel NOMA assisted IoVT framework with multi-MEC is proposed and optimization is performed in the proposed framework, followed by performance evaluation and analysis. Finally, we conclude this article and discuss future research issues.

\section*{\large{Challenges of IoVT Networks with Multi-MEC}}

\begin{figure}[htb]
	\centering
	\includegraphics[width=7.5cm]{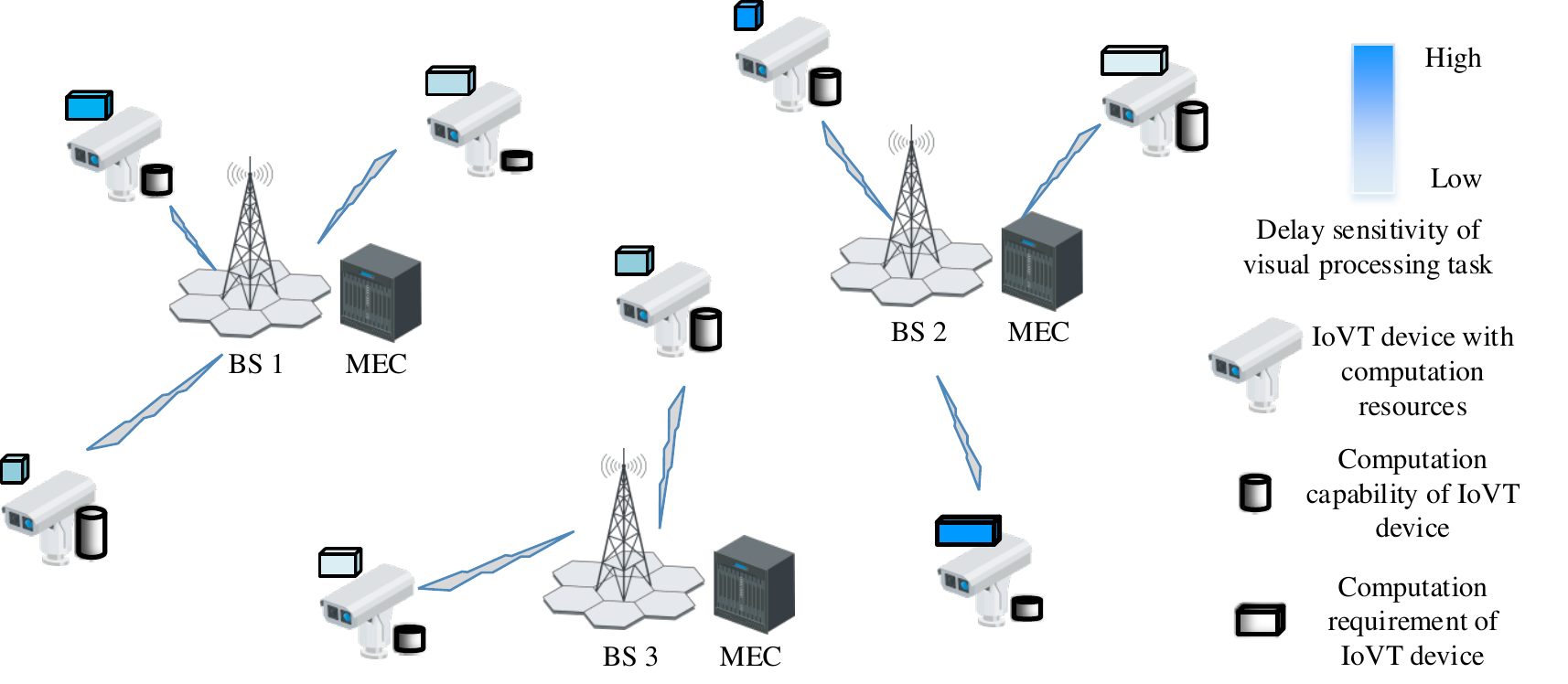}
	\caption{A heterogeneous IoVT system with multi-MEC.}\label{figure:Scenario}
\end{figure}

An IoVT scenario with multiple BSs is illustrated in Fig.\ref{figure:Scenario}, where each BS is deployed with an MEC server. To show the heterogeneity of the IoVT network, different IoVT devices have different computation capability and request different visual processing tasks. The computation demands and delay sensitive of the visual processing tasks are different, and are indicated by different lengthes and different colors of the rectangles beside IoVT devices, respectively. Computation capability of different IoVT devices are indicated by different heights of the cylinders beside IoVT devices. To meet the computation and delay demands of the visual processing tasks, IoVT devices offload partial visual processing tasks to MEC servers located at BSs.

In order to reduce the total delay as well as to make full use of computation resources at multiple MEC servers and IoVT devices in the heterogeneous IoVT network, there are several challenges to be addressed. These challenges can be summarized as issues on MEC association, visual data uplinking and task division, corresponding to three problems (i.e., where to offload? how to offload? and which to be offloaded?).

{\bf MEC Association:} To meet the huge computation requirements from visual processing, multiple MEC servers are deployed with BSs. In such multi-MEC IoVT network, the first problem for each IoVT device is where the visual processing tasks should be offloaded.

As each MEC server is co-located with an BS, the MEC association problem is equivalent to the user association problem in wireless networks. Traditional association strategies are based on channel conditions, where a user associates to an BS with the nearest distance or the strongest signal strength. However, in the context of multi-MEC offloading, these association strategies might result in unbalanced computation load among MEC servers and thus increase the possibility of offloading failures. For example, for an IoVT device, the MEC server co-located with its nearest BS might not has sufficient computation resources to satisfy its visual processing task offloading demand, while some other MEC servers within its association range has enough computation resources. Therefore, to achieve better utilization efficiency of computation resources for MEC servers and reduce offloading failures, besides channel conditions, visual processing task offloading demands of different IoVT devices and computation resources at different MEC servers should also be jointly considered in designing the association strategy for multi-MEC offloading.

{\bf Visual Data Uplinking:} In the IoVT network, there is a contradiction between limited wireless uplink throughput and huge visual data transmission requirements. Hence, the problem of how to offload required visual data to MEC servers over wireless links is critical for visual processing task offloading, which is also referred to as visual data uplinking problem.

Most existing work focuses on compression based uplinking technologies where visual data is compressed before offloaded to MEC servers. However, visual data compression will incur additional computation resources and power consumption at the IoVT devices. Another way to address the visual data uplinking problem is to increase uplink throughput by improving the resource utilization efficiency of wireless links. In this way, one of the most promising technologies is NOMA. With NOMA transmission, multiple signals are superposed and transmitted simultaneously using the same time/frequency resource.
{There have been some works that combine NOMA with MEC in the traditional wireless network to reducing total delay or total energy consumption.
To minimize the overall delay of all users, in \cite{Wu2018Noma}, the authors jointly optimize the workloads division and time domain resource allocation of users in a NOMA-assisted multi-access MEC offloading system. In this system, the computation task of each user is divided into several parts and these parts are offloaded to different MEC servers.}
Nevertheless, the coupling relationship with MEC association and delay constraints on offloaded visual processing tasks make the application of NOMA in the IoVT network is much more challenging than in the traditional wireless network. Particularly, more sophisticated power allocation is necessary for visual data uplinking according to the energy constraints at the IoVT devices.

{\bf Task Division:} In the IoVT network, both IoVT devices and MEC servers have computation capability for visual processing. In this case, the problem that which visual processing tasks should be chosen to be offloaded to the MEC servers arises. Each IoVT device will divide the visual processing tasks into two parts, one for local processing and the other for MEC offloading. The results of task division will directly determine the MEC offloading performance in terms of delay.

In essence, task division is to match the visual processing tasks with computation capability at the MEC servers and IoVT devices under the delay constraint on each visual processing task, aiming at minimizing the total delay of all visual processing tasks. The heterogeneity of the visual processing task offloading demands from the IoVT devices as well as the heterogeneity of the computation capability of the MEC servers and IoVT devices make it intractable to achieve optimal task division. For each IoVT device, computation capability of the MEC server that it associates to is an important factor that should be considered in task division. In the multi-MEC IoVT network, there might be multiple MEC servers in the association range of the IoVT device. Hence, The task division problem is much more complex for multi-MEC offloading.

\begin{figure*}[htb]
	\centering
	\includegraphics[angle=0,scale=0.66]{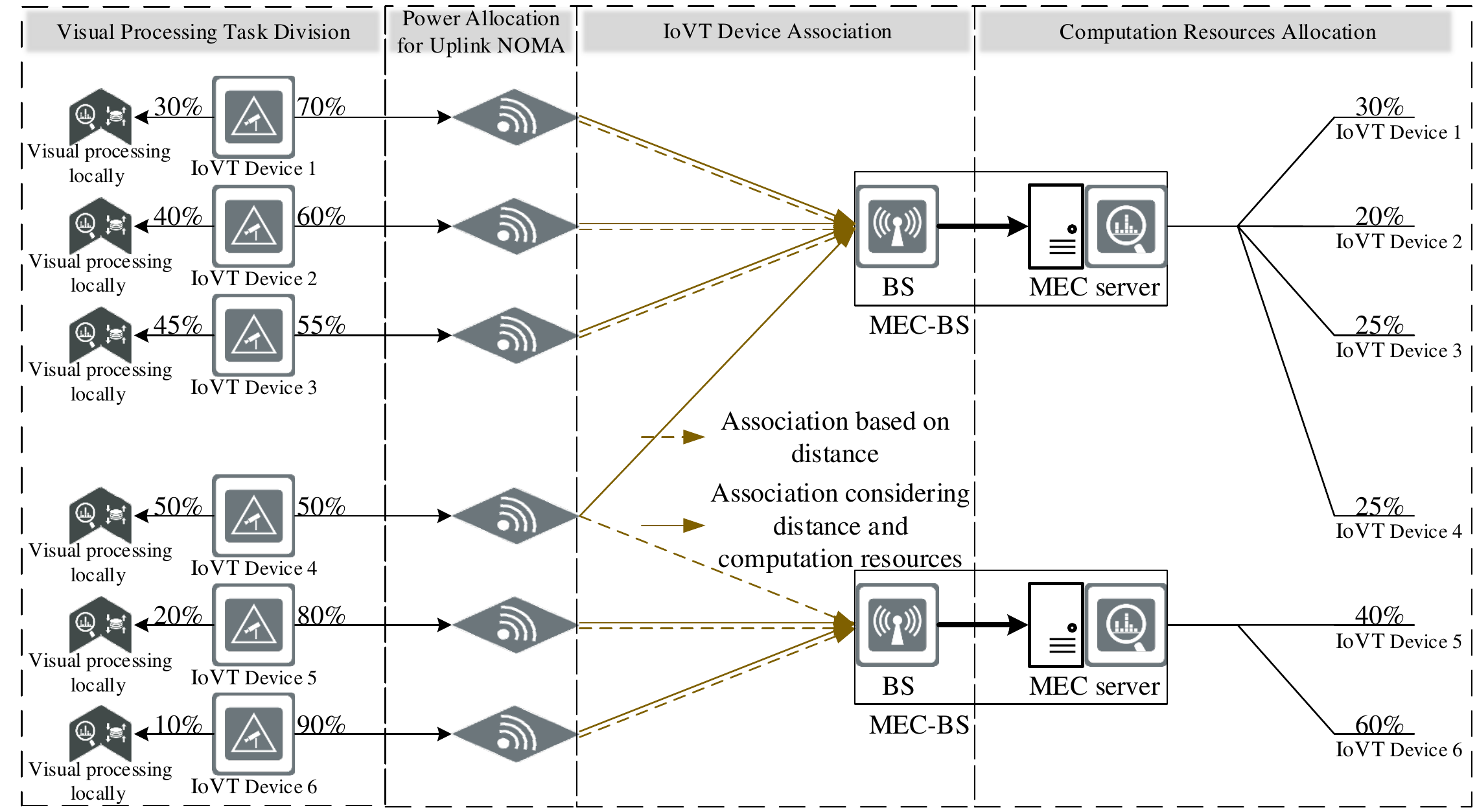}
	\caption{The proposed NOMA assisted IoVT framework with multi-MEC.}\label{fig:network}
\end{figure*}

\section*{\large{A Novel NOMA Assisted IoVT Framework with Multi-MEC}}

To address the challenges in the previous section, we propose a novel NOMA assisted IoVT framework with multi-MEC as shown in Fig.\ref{fig:network}. In the proposed framework, we jointly optimize the association strategy, uplink wireless transmission assisted by NOMA and division of visual processing tasks as well as computation resources to meet the delay requirements of all IoVT devices, with the goal to minimize the total delay of all visual processing tasks. The work flow of the proposed joint optimization are described in Fig.\ref{figure:Framework_and_NOMA}.

\subsection*{\large{Game Based MEC Association}}
In this article, we consider an IoVT network where an MEC server is deployed with each BS. For convenience, an MEC server and the BS with which the MEC server located is denoted as an MEC-BS. We assume that there are $M$ MEC-BSs and $N$ IoVT devices in the IoVT network. With consideration of visual processing task offloading demands of IoVT devices and computation capability of MEC-BSs, we model the MEC association problem as a many-to-one matching game where IoVT devices and MEC-BSs are two player sets, denoted by $\mathcal{V}=\{v_1,v_2,\ldots,v_N\}$ and set $\mathcal{B}=\{b_1,b_2,\ldots, b_M\}$, to be matched together.

We define a matching strategy $\Phi$ from set $\mathcal{V}$ to set $\mathcal{B}$. $\Phi(v_i)=b_j$ means that IoVT device $v_i$ associates to MEC-BS $b_j$. In particular, $\Phi(v_i)=\varnothing$ means that IoVT device $v_i$ does not associate to any MEC-BS. In the matching game, each player have its own preference list. That is to say, an IoVT device prefers to associate to the MEC-BS with better channel conditions and sufficient computation resources, while an MEC-BS prefers to serve more IoVT devices with better channel conditions under delay constraints of visual processing tasks. The preference lists of $v_i$ and $b_j$ are denoted by $\mathcal{P}(v_i)$ and $\mathcal{P}(b_j)$, respectively. The utility functions of $v_i$ and $b_j$ with matching strategy $\Phi$ are denoted by $l(\Phi,v_i)=\mathbf{I}(\Phi(v_i) \neq \varnothing)$ and $l(\Phi,b_j)=\sum\limits_{v_i \in \mathcal{V}} \mathbf{I}(\Phi(v_i)= b_j)$, respectively, where $\mathbf{I}(\cdot)$ is the indicator function. Let $\Gamma (\mathcal{V}, \mathcal{B}, \Phi)=\sum\limits_{v_i \in \mathcal{V}} \mathbf{I}(\Phi(v_i) \neq \varnothing)$ denote the number of IoVT devices associating to MEC-BSs when matching strategy $\Phi$ is applied. We define the many-to-one matching game as the following tuple.

\begin{equation}\label{3622}%\nonumber
\setlength\abovedisplayskip{1pt}
	\setlength\belowdisplayskip{1pt}
\mathscr{G}=\Big(\mathcal{V}, \mathcal{B}, \Phi, \mathcal{P},\Gamma (\mathcal{V}, \mathcal{B}, \Phi)\Big),
\end{equation}

{The Nash Equilibrium (NE) is a strategy profile where no player has incentive to deviate unilaterally.
The definition of NE of the proposed game model is as follows:

\newtheorem{def1}{\bf Definition}
{\begin{def1}: An NE of game $\mathscr{G}$ is reached if no IoVT device can associated to a preferred MEC-BS and no MEC-BS can accept more IoVT devices.\end{def1}}}

To find an NE, Gale-Shapley algorithm is adopted, which is described as follows.

1) Each IoVT device (e.g., $v_i,~(1\leq i\leq N)$), sends an association request to the first MEC-BS (e.g., $b_j,~1\leq j \leq M$), in its preference list $\mathcal{P}(v_i)$ (i.e., $\Phi(v_i)=b_j$);

2) Each MEC-BS (e.g., $b_j,~(1\leq j \leq M)$), attempts to serve IoVT devices associating to it. If the delay and computation resource requirements of all IoVT devices can be satisfied, $b_j$ accepts all of them. Otherwise, $b_j$ rejects the last IoVT device (e.g., $v_i$) that associates to it according to its preference list $\mathcal{P}(b_j)$, and set $\Phi(v_i)=\varnothing$. %(This step will repeat until all BSs can satisfy the delay requirements and computation demands of all IoVT devices being accepted.)

3) If IoVT device $v_i$ is rejected by MEC-BS $b_j$, $v_i$ will delete $b_j$ from its preference list (i.e., $\mathcal{P}(v_i)=\mathcal{P}(v_i)\setminus b_j$), and sends an association request to the first MEC-BS (e.g., $b_k$) in its new preference list $\mathcal{P}(v_i)$ (i.e., $\Phi(v_i)=b_j$);

4) Steps 2-3 will repeat until all IoVT devices finish their association or all MEC-BS can not accept any more association request (i.e., $\mathcal{P}(v_i)=\varnothing$ if $\Phi^*(v_i) \neq \varnothing$).

{Based on this algorithm, we can prove that Game $\mathscr{G}$ is bound to stop at a NE after a limited number of iterations: In this algorithm, each IoVT sends association requests to MEC-BSs in the descending order of its preference list. Therefore, when game $\mathscr{G}$ stop, no IoVT player can associated to a preferred MEC-BS, and no MEC-BS can accept more IoVT devices. In addition, the size of the preference list of unassociated V-IoT player $v_i$, $\mathcal{P}(v_i)$, is reduced after each iteration. In other words, a matching strategy $\Phi$ will not be selected repeatedly. So game $\mathscr{G}$ is bound to stop at a NE after a limited number of iterations.}

\subsection*{\large{NOMA Based Visual Data Uplinking}}
After MEC association, uplink NOMA transmission is performed by the IoVT devices associating to the same MEC-BS. Consider an MEC-BS $b_j$. Let $V(b_j)$ denote the set of IoVT devices that associate to $b_j$. All IoVT devices simultaneously transmit the signals that carry their visual data information. In this case, $b_j$ receives a superposition form of the signals from these devices. With SIC, $b_j$ first decodes the signal with the strongest strength for the IoVT device that transmits it by regarding the signals from other IoVT devices as noise, and then subtracts the decoded signal from the original received signal. By repeating this process, all signals from different IoVT devices can be decoded.

\begin{figure*}[htb]
	\centering
	\includegraphics[width=16cm]{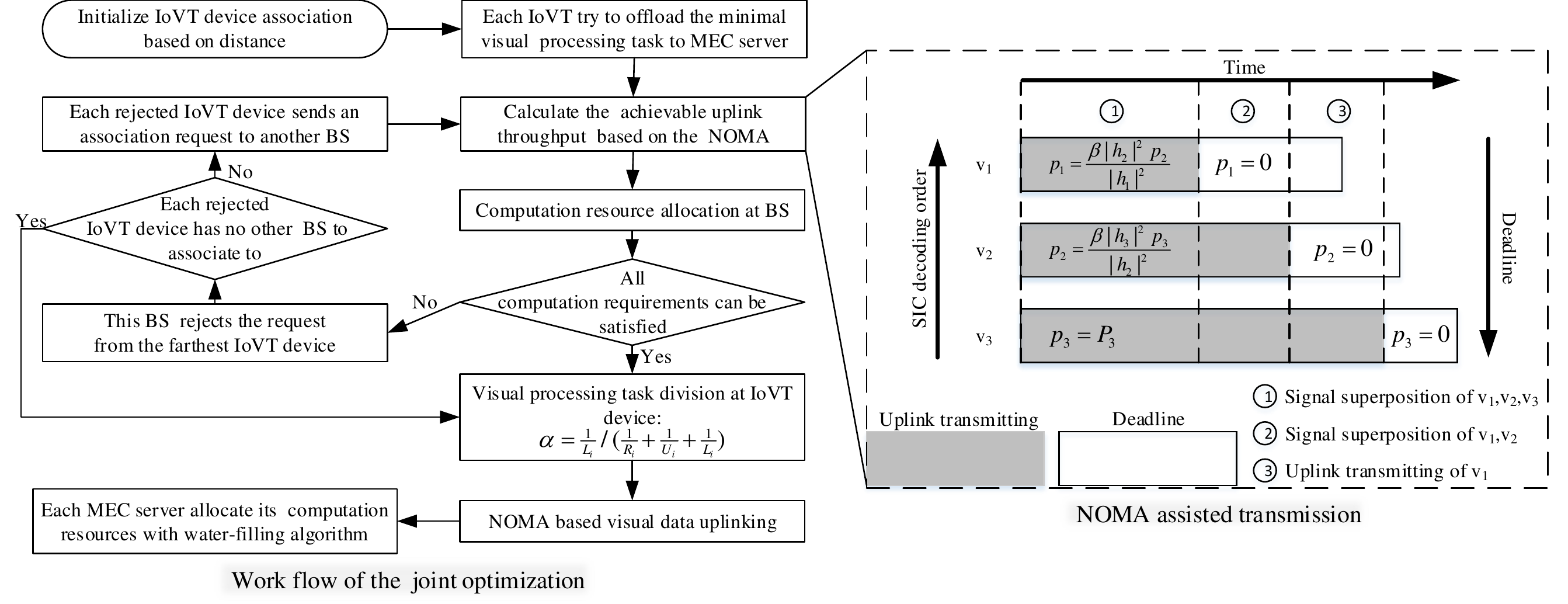}
	\caption{The work flow of proposed joint optimization and NOMA assisted transmission.}\label{figure:Framework_and_NOMA}
\end{figure*}

The SIC decoding order determines the offloading delay of the IoVT devices. The IoVT device whose visual data are decoded first will suffer low offloading delay. In the traditional uplink NOMA system, the SIC decoding order is the descending order of channel conditions. However, in the context of MEC offloading, different IoVT devices have different delay requirements for visual processing task offloading. Hence, we propose a deadline aware SIC decoding order, which is described in the right of Fig.\ref{figure:Framework_and_NOMA}. Three IoVT devices transmit their signals simultaneously in stage $\textcircled{\small{\small{1}}}$ until uplink transmission of $v_1$ is finished. The MEC-BS decodes the signals of these IoVT devices by SIC in ascending order of their delay requirements (i.e., deadlines). The signal of $v_1$ is first decoded. It also means that $v_1$ is the first one to finish its uplink transmission. After stage $\textcircled{\small{\small{1}}}$, $v_2$ and $v_3$ transmit their signals simultaneously in stage $\textcircled{\small{\small{2}}}$ until uplink transmission of $v_2$ is finished. Then $v_3$ transmits its signal in stage $\textcircled{\small{\small{3}}}$.

To achieve the desired SIC decoding order, power allocation should be carefully considered at the IoVT devices based on their channel conditions. Assume that transmit power of two IoVT devices $v_i$ and $v_k$ are $p_i$ and $p_k$, respectively, and the channel gains are $h_i$ and $h_k$, respectively \cite{WangYazheng2021L}. If visual data of $v_k$ is decoded before that of $v_i$, $|h_{i}|^2p_{i}<\beta|h_{k}|^2p_{k}$ should be satisfied, where $\beta\in(0,~1]$ can be set according to decoding capability of MEC-BS $b_j$. Based the above analysis, we can design the power allocation strategy to satisfy the delay requirements of the IoVT devices. An example is shown in the right of Fig.\ref{figure:Framework_and_NOMA}, where transmit power of $v_i$ is allocated as $p_{i}=min(\tfrac{\beta|h_{i-1}|^2p_{i-1}}{|h_{i}|^2},P_i),~ i=1, 2, 3$. $P_i$ is the transmit power constraint of $v_i$.

\subsection*{\large{Delay Driven Visual Processing Task Division}}
{For each IoVT device, the visual processing task of each IoVT device is divided into two parts, one part is processed locally and the other one part is processed at MEC server. We assume that computation capability can be indicated by the visual processing rate.
For IoVT device $v_i$,  let $L_i$, $R_i$ and $U_i$ denote its local visual processing rate, achievable uplink throughput and the visual processing rate that the MEC server allocated to it, respectively. Let $C_i$ and $\alpha_i$ denote the total computation-workload requirement of IoVT device $v_i$ and the proportion of its visual processing task offloaded to the MEC server, respectively. Then the local visual processing delay, the delay produced by visual data offloading and visual processing at the MEC server are $\tfrac{(1-\alpha)C_i}{L_i}$, $\tfrac{\alpha C_i}{R_i}$ and $\tfrac{\alpha C_i}{U_i}$, respectively.

The final delay of each visual processing task is determined by the larger of the local visual processing delay and the delay produced by visual data offloading and visual processing at the MEC server, i.e. $max(\tfrac{(1-\alpha)C_i}{L_i}, \tfrac{\alpha C_i}{R_i}+\tfrac{\alpha C_i}{U_i})$. To make full use of both the local computation resources and the computation resources at the MEC server, task division should make these two kinds of delay be the same, i.e. $\tfrac{(1-\alpha)C_i}{L_i}=\tfrac{\alpha C_i}{R_i}+\tfrac{\alpha C_i}{U_i}$. Then, we can obtain optimal task division as $\alpha=\tfrac{1}{L_i}/(\tfrac{1}{R_i}+\tfrac{1}{U_i}+\tfrac{1}{L_i})$.}

Each MEC server performs computation resource allocation to handle the visual processing tasks offloaded by the IoVT devices associating to it. Two principles are considered by the MEC server: 1) satisfy as many computation resource requirements of the offloaded visual processing tasks as possible. 2) minimize the total delay of all visual processing tasks. Based on these two principles, the MEC server first allocates the minimum required computation resources to each offloaded visual processing task. Then, the remaining computation resources are allocated to the offloaded visual processing task whose delay can be reduced most. Actually, such computation resource allocation problem can be treated as a water-filling problem, which has been well studied in wireless communication.

{In the proposed framework, the issues on MEC association, visual data uplinking and task division are related to each other because of delay. If an IoVT device associate to the nearest BS-MEC where the computation
resources are insufficient, although the delay for visual data uplinking may be very small, the delay produced by visual processing at the MEC server will be too large to meet the deadline.
Hence as shown in Fig.3, in the work flow of the proposed framework, each IoVT device first try to associate to the neatest BS-MEC. Then the achievable uplink throughput based on NOMA (i.e., $R_i$) is calculated. If this MEC server can provide sufficient computation resources to this IoVT device, so that the visual processing tasks can be finished before deadline, this BS-MEC will accept the association request of this IoVT device, otherwise this BS-MEC will reject this association request and this IoVT device will try to associate to another BS-MEC. After the IoVT association is finished, each MEC server will divide its computation resources to the associated IoVT devices by water-filling algorithm.}

\section*{\large{Performance evaluation}}

\begin{figure}
    \centering
\subfigure[]{
\label{fig:dd}
\includegraphics[width=6.5cm]{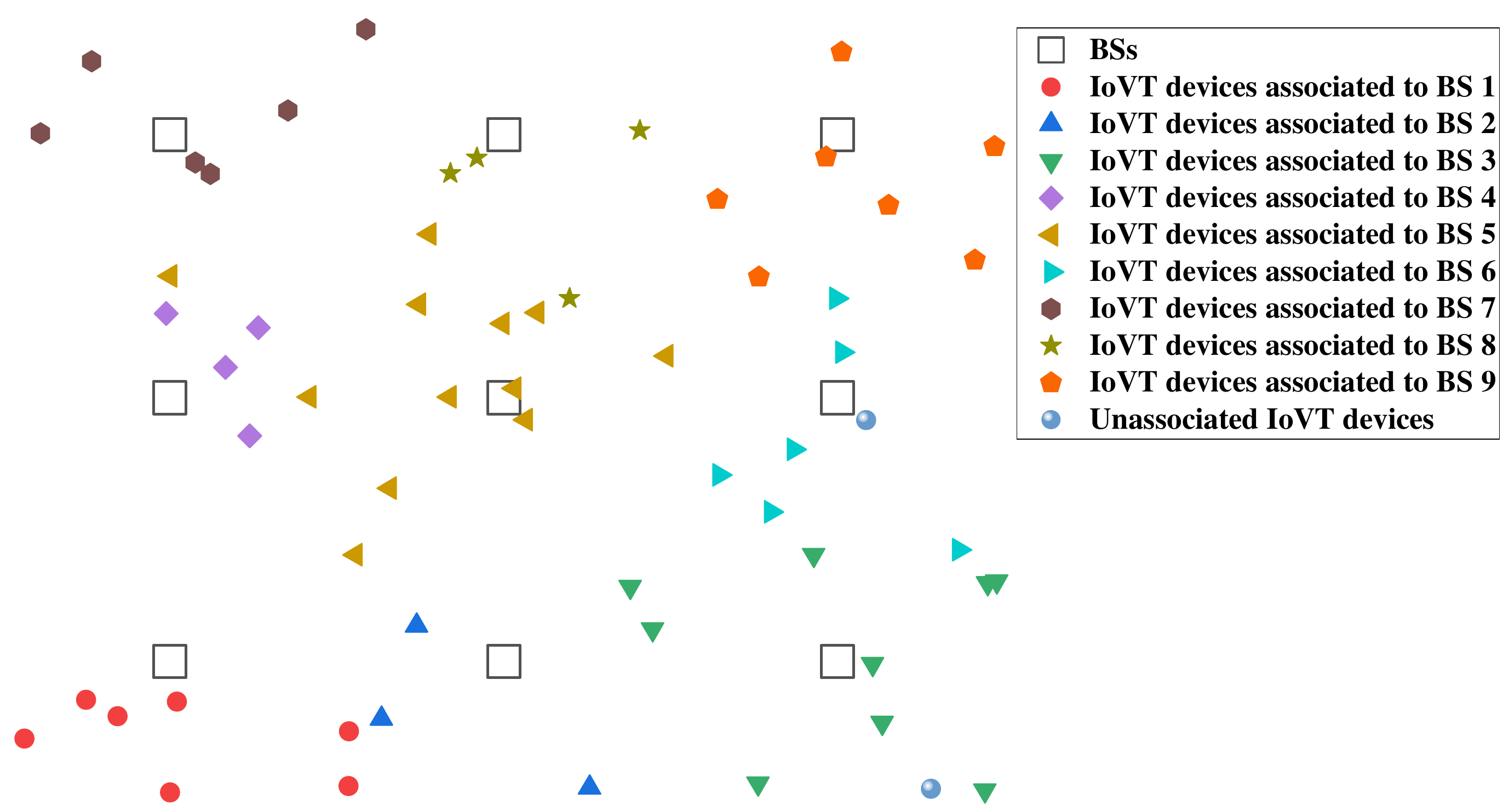}}
\subfigure[]{
\label{fig:dd}
\includegraphics[width=6.5cm]{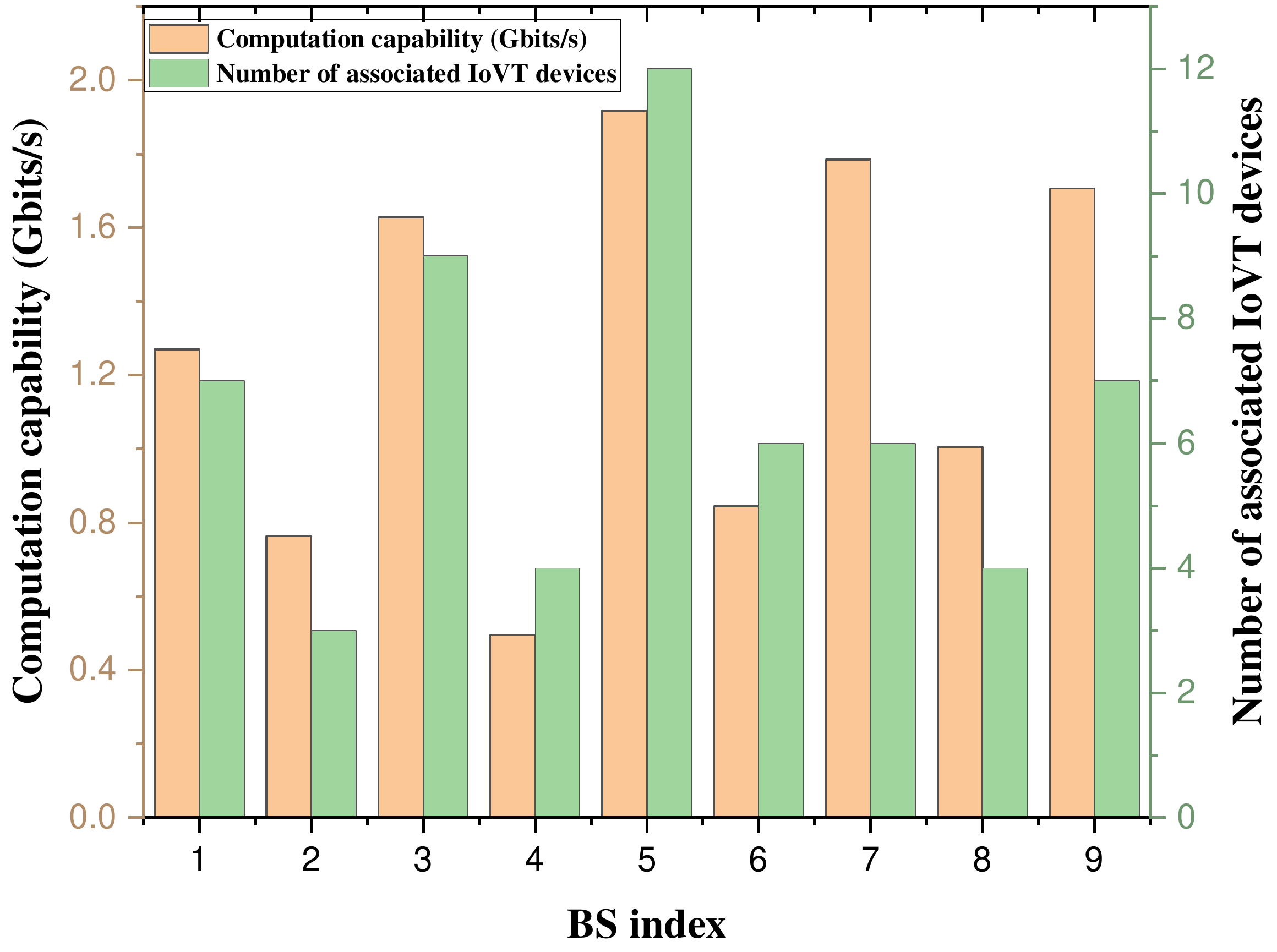}}
    \caption{{Simulation results. (a) {Distribution of BSs and IoVT devices} (b) Total delay vs. Number of IoVT devices.}}
    \label{simuResults2}
\end{figure}

In this section, the performance of the proposed framework is evaluated by simulations. In the simulations, nine BSs are respectively placed in (-200m, 200m), (0m, 200m), (200m, 200m), (-200m, 0m), (0m, 0m), (200m, 0m), (-200m, -200m), (0m, -200m), (200m, -200m) and IoVT devices are randomly distributed in a 600m$\times$600m rectangular area around BSs, as shown in Fig. \ref{simuResults2}(a). Each BS is deployed with an MEC server.
The large-scale channel gain is $128.1\!+\!37.6$log$_{10}( d_{n}[$km$])$ dB.
The Rayleigh fading coefficient follows an i.i.d. Gaussian distribution as $\beta\!\!\thicksim\!\mathcal{CN}(0, 1)$. We set the noise power to $\sigma^2=BN_0$, where $B\!\!= \!\!2$ MHz and $N_0\!= $-$174$ dBm/Hz. The computation-workload requirement of each IoVT device is randomly chosen from $5$ to $10$ Mbits. The deadline of each IoVT device is randomly chosen from $0.1$ to $2$ s. Computation capability of each IoVT device is randomly chosen from $1$ to $10$ Mbits/s. Computation capability of each MEC server is randomly chosen from $0.4$ to $2$ Gbits/s.

{In this paper, if a BS-MEC can-not provide enough computation Resources to an IoVT device, the BS-MEC will reject the association request of this IoVT device. If an IoVT device is rejected by all BS-MEC, we call it ``Unassociated IoVT devices'' as shown in Fig. \ref{simuResults2}(a).} Fig. \ref{simuResults2}(a) shows the distribution of $9$ BSs and $60$ IoVT devices in the proposed framework. The IoVT devices in the same color associate to the same BS. The distance between the IoVT device and BS indicates channel condition.
Fig. \ref{simuResults2}(b) shows computation capability and the number of IoVT devices associating to each BS.

{In this paper, each MEC server is co-located with an BS, so the IoVT devices not only prefer to associate with the BS with the best channel condition but also the BS with the MEC server with sufficient computation resources. Furthermore, considering the uneven distribution of the IoVT devices, if all IoVT devices associate to the nearest BS, the computation load among MEC servers may be unbalanced and thus the possibility of offloading failures increases. Therefore, the traditional association strategies according to distance are no longer effective. In the proposed framework, the channel condition and the computation resources are jointly considered in MEC association as stated in the game process of matching game $\mathscr{G}$. So in Fig. \ref{simuResults2}(a), when the computation resources of the nearest MEC server is insufficient, the IoVT devices may associate to the other MEC server.} By considering both computation capability and channel conditions when performing IoVT device association, the MEC server with more computation resources will serve more IoVT devices as shown in Fig. \ref{simuResults2}(b). This result confirms that the proposed association strategy can allocate computation resources of MEC servers more reasonably.

\begin{figure}
    \centering
\subfigure[]{
\label{fig:dd}
\includegraphics[width=6.5cm]{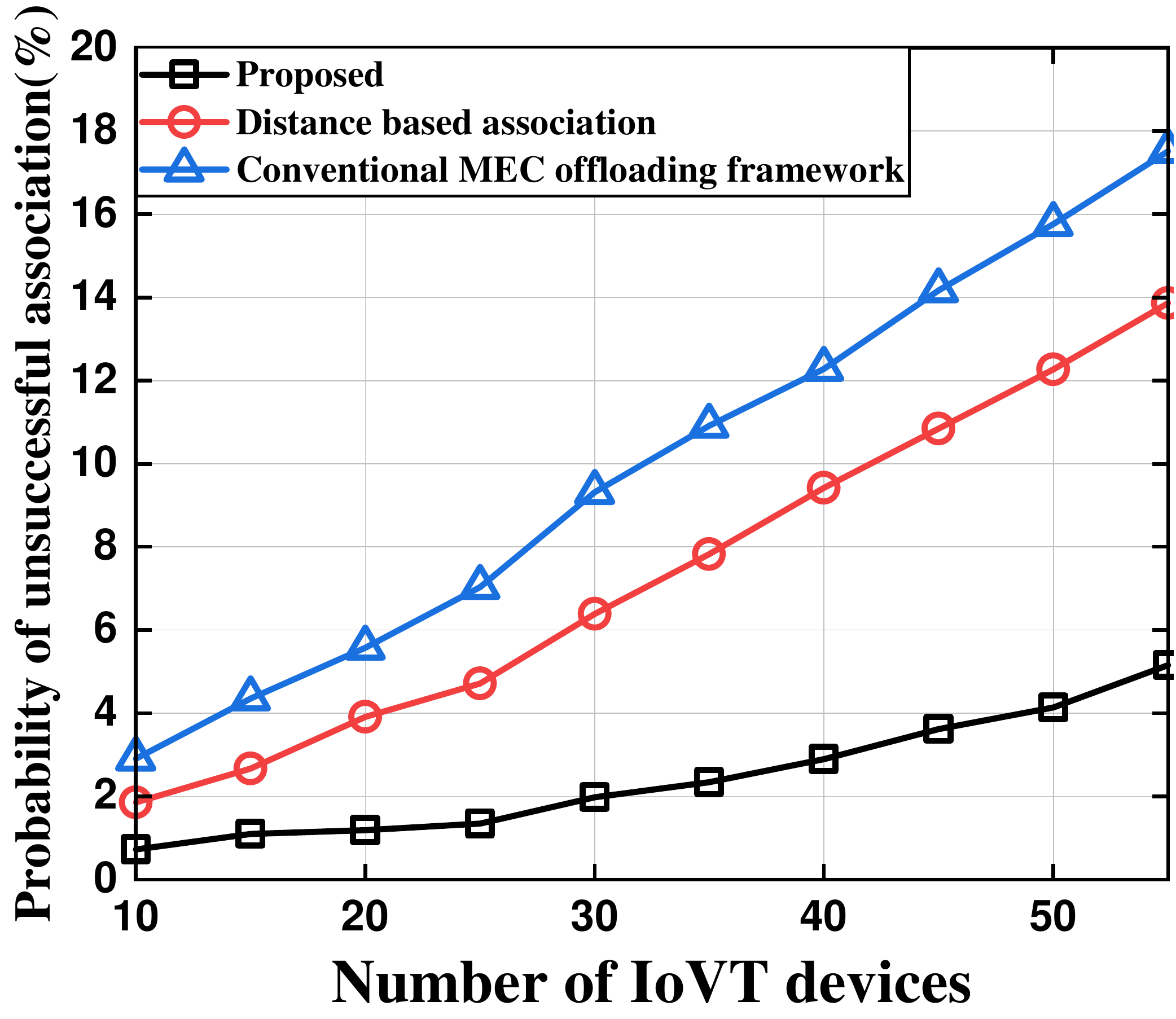}}
\subfigure[]{
\label{fig:dd}
\includegraphics[width=6.5cm]{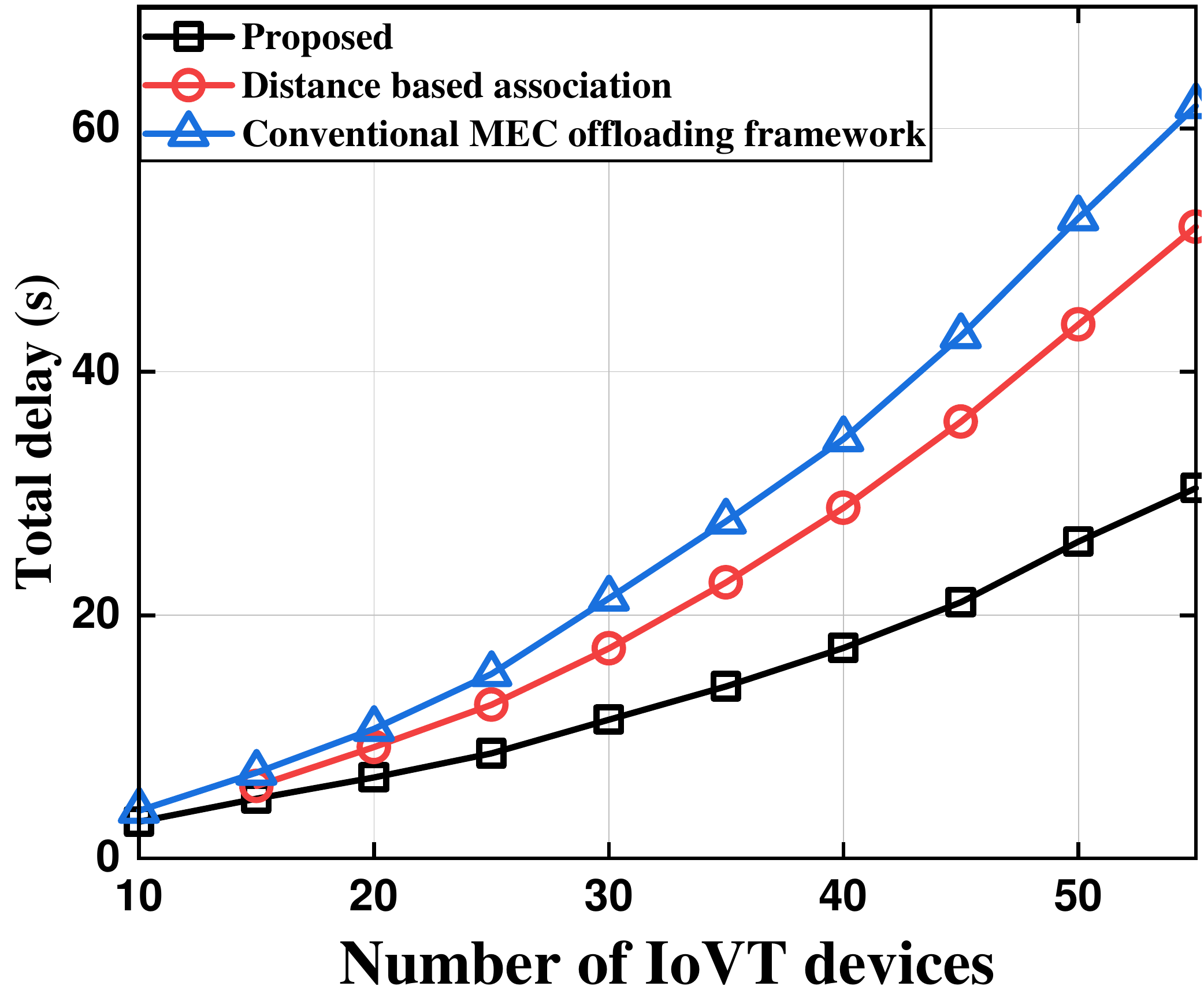}}
    \caption{{Simulation results. (a) Probability of successful association vs. Number of devices (b) Total delay vs. Number of devices.}}
    \label{simuResults}
\end{figure}

To evaluate the performance of proposed joint optimization, in Fig. \ref{simuResults}(a) and Fig. \ref{simuResults}(b), we vary the number of IoVT devices from 10 to 55 to observe its effect on the probability of unsuccessful association and the total delay of all IoVT devices, respectively.
The delay of the IoVT device without association is set to $10$s for penalty.
For comparison, the red lines are the results of the proposed framework with the traditional distance based association,
and the blue lines are the conventional MEC offloading framework where the traditional distance based association and the channel condition based SIC decoding order are adopted.
We can see that both the probability of unsuccessful association and the total delay of all IoVT devices increase with the number of IoVT devices. The reason is that, as the number of IoVT devices increases, the interference among IoVT devices associating to the same BS will increase, which will lead to lower uplink throughput. Therefore, the transmission delay will increase. Furthermore, considering that the computation resources of each MEC server is limited, the computation resources allocated to each IoVT device will be reduced as the number of IoVT devices increases. In addition, proposed joint optimization outperforms the reference schemes. The reason can be described as follows. First, the visual processing task with nearest deadline will be transmitted with the highest priority by NOMA at the IoVT device, therefore, the probability of unsuccessful association is reduced. Second, with the proposed association strategy, the MEC server with more computation resources will serve more IoVT devices as shown in Fig. \ref{simuResults2}(b). In other words, proposed computation resource allocation is more reasonable.

\section*{\large{Conclusion and Future Work}}

To break through the bottlenecks of uplink throughput and computation capability of single MEC server in IoVT networks, in this article, we propose a novel NOMA assisted IoVT framework with multi-MEC. In the proposed framework, to minimize the total delay of all visual processing tasks, we perform joint optimization on the association strategy, uplink wireless transmission assisted by NOMA, division of the visual processing tasks at the IoVT devices and computation resource allocation at the MEC servers under the delay constraint of each IoVT device. Simulation results show that the total delay and the probability of unsuccessful association of the IoVT device can be reduced significantly compared with the reference schemes.

As future work, some open issues on MEC offloading in IoVT networks are discussed as follows.
%{\bf BS Deployment:}
%Considering that the locations of IoVT devices is fixed, when deploying BSs to server these devices, the locations of BSs will affect the channel condition between IoVT devices and BSs and then influence
%the result of association. In the future, we will focus on how to deploy BSs to server more IoVT devices with lower deployment cost.

{\bf Cooperative Offloading with Cloud:}
Due to powerful computation capability of cloud, many studies propose to offload the visual processing tasks to cloud. In this case, task division of visual processing tasks will become more challenging with additional consideration of computation capacity of cloud. Visual processing task offloading should be cooperatively performed by the IoVT devices, MEC servers and cloud servers. Furthermore, as cloud servers are usually serval hops away from the IoVT devices, the visual data uplinking problem will much more complex. New research issues, such as backhaul scheduling, should be concerned.

{\bf Evaluation Metrics for Offloading:}
More sophisticated evaluation metrics besides delay are expected according to IoVT applications. For example, the evaluation metrics for human oriented visual communication is much different from that for machine-to-machine visual communication. For the former, perceptual visual quality is important, while the latter prefers to transform complex visual scenes into simple words. Evaluation metrics have significant impact on transmission, processing and offloading of visual data. The MEC offloading mechanism should be carefully designed according to the evaluation metrics.

{\bf Joint Visual Data Coding and Uplinking:}
As IoVT devices become more and more intensive in smart cities, especially in monitoring for public security, it is likely that there exists quite a lot of redundancy among the visual data generated by neighboring IoVT devices. The redundancy will result in a huge waste of uplink transmission and computation resources. To deal with this issue, collaborative coding and wireless transmission can be jointly studied. For example, distributed video coding and NOMA uplink transmission can be integrated to achieve optimal offloading performance.

\section*{Acknowledgment}

%This work was supported in part by the National Science Foundation of China (NSFC) (Grants 91538203, 61631017, 61390513 and 61771445),
%and the Fundamental Research Funds for the Central Universities.

This work was supported by National Key R\&D Program of China under Grant 2020YFA0711400 and National Science Foundation of China under Grant 61771445, 61631017, 91538203.

\bibliographystyle{ieeetr}

\section*{Biographies}

\noindent
\footnotesize{{Fengqian Guo} received the B.S. degree in 2017 from the Jiangnan University, Wuxi, China, where he is currently pursuing the Ph.D. degree in communication and information systems with the Department of Electronic Engineering and Information Science, University of Science and Technology of China (USTC), Hefei, China. His current research interests include wireless transmission and multiple access networks.}\\

\noindent
\footnotesize{{\sc Hancheng Lu}
(M'07) received the Ph.D. degree in communication and information systems from the University of Science and Technology of China (USTC), Hefei, China, in 2005. He is currently an associate professor with the Department of Electronic Engineering and Information Science, USTC. His research interests include multimedia communication and networking, resource optimization in wireless heterogeneous networks.}\\

\noindent
\footnotesize{{\sc Bo Li} received the B.S. degree in 2020 from the Dalian University of Technology (DLUT), Dalian, China. He is currently pursuing the M.S. degree in electronics and communication engineering with the Department of Electronic Engineering and Information Science, University of Science and Technology of China(USTC), Hefei, China. His current research interests include wireless communication.}\\

\noindent
\footnotesize{{\sc Dingxuan Li} received the B.E. degree in 2020 from the University of Science and Technology of China (USTC), Hefei, China, where he is currently working toward the MA.Eng. degree in communication and information systems with the Department of Electronic Engineering and Information Science. His research interests include wireless network power allocation and federated learning.}\\

\noindent
\footnotesize{{\sc Chang Wen Chen} (F'04) is currently Chair Professor of Visual Computing at The Hong Kong Polytechnic University. He has previously served as an Empire Innovation Professor of Computer Science and Engineering at the University at Buffalo, State University of New York from 2008 to 2021. From 2017 to 2020, He served as Dean of the School of Science and Engineering at The Chinese University of Hong Kong, Shenzhen. He was Allen Henry Endow Chair Professor at the Florida Institute of Technology from 2003 to 2007. He was on the faculty of Electrical and Computer Engineering at the University of Rochester from 1992 to 1996 and on the faculty of Electrical and Computer Engineering at the University of Missouri-Columbia from 1996 to 2003.

He has been the Editor-in-Chief for IEEE Trans. Multimedia from 2014 to 2016. He has also served as the Editor-in-Chief for IEEE Trans. Circuits and Systems for Video Technology from 2006 to 2009. He has been an Editor for several other major IEEE Transactions and Journals, including the Proceedings of IEEE and IEEE Journal of Selected Areas in Communications. He has served as Conference Chair for several major IEEE, ACM and SPIE conferences related to multimedia, video communications and signal processing, including ACM MM2020 and ACM MM2018 in recent years. His research expands a broad range of topics in multimedia communication, Internet of Video Things,  multimedia systems, image/video processing, machine learning, and multimedia signal processing.

He received his BS from University of Science and Technology of China in 1983, MSEE from University of Southern California in 1986, and Ph.D. from University of Illinois at Urbana-Champaign in 1992.  He and his students have received nine (9) Best Paper Awards or Best Student Paper Awards. He has also received several research and professional achievement awards, including Sigma Xi Excellence in Graduate Research Mentoring Award in 2003, Alexander von Humboldt Research Award in 2009, the University at Buffalo Exceptional Scholar Sustained Achievement Award in 2012, the State University of New York System Chancellor's Award for Excellence in Scholarship and Creative Activities in 2016, and the Distinguished ECE Alumni Award from University of Illinois at Urbana-Champaign in 2019. He is an IEEE Fellow since 2005 and an SPIE Fellow since 2007.}\\

%\section*{\large{Biographies}}

%\bibliography{mybib}
\end{document}